\documentclass[final,5p,times,twocolumn]{elsarticle}
\usepackage[T1]{fontenc}
\usepackage[utf8]{inputenc}

\usepackage{amsmath}
\usepackage{amsfonts}
\usepackage{amssymb}
\usepackage{amsxtra}
\usepackage{array}
\usepackage{color}
\usepackage{dcolumn}
\usepackage{graphicx}
\usepackage{hepunits}
\usepackage{xspace}
\usepackage{CJKutf8}

\usepackage{subcaption}

\definecolor{purple}{rgb}{0.5,0,0.5}
\definecolor{blue}{rgb}{0.0,0,0.9}
\definecolor{prdblue}{rgb}{0.133,0.118,0.498}
\usepackage[colorlinks=true, pdfstartview=FitV, linkcolor=prdblue, citecolor= prdblue, urlcolor=prdblue]{hyperref}

\usepackage[mathscr,scaled=1.15]{urwchancal}
\DeclareFontFamily{OT1}{pzc}{}
\DeclareFontShape{OT1}{pzc}{m}{it}%
{<-> s * [1.15] pzcmi7t}{}
\DeclareMathAlphabet{\mathpzc}{OT1}{pzc}{m}{it}

\biboptions{sort&compress}

\journal{Physics Letters B}

\newcommand{\beq}{\begin{equation}}
\newcommand{\eeq}{\end{equation}}
\newcommand{\ba}{\begin{array}}
\newcommand{\ea}{\end{array}}
\newcommand{\bea}{\begin{align}}
\newcommand{\eea}{\end{align}}
\newcommand{\bi}{\begin{itemize}}
\newcommand{\ei}{\end{itemize}}
\newcommand{\ben}{\begin{enumerate}}
\newcommand{\een}{\end{enumerate}}
\newcommand{\bc}{\begin{center}}
\newcommand{\ec}{\end{center}}
\newcommand{\bl}{\begin{flushleft}}
\newcommand{\el}{\end{flushleft}}
\newcommand{\br}{\begin{flushright}}
\newcommand{\er}{\end{flushright}}

\begin{document}
\begin{CJK*}{UTF8}{gbsn}

\begin{frontmatter}

\title{$\,$\\[-7ex]\hspace*{\fill}{\normalsize{\sf\emph{Preprint no}. NJU-INP 118/26}}\\[1ex]
Distribution amplitudes and functions of ground-state scalar and pseudoscalar charmonia}

\author[NJU,INP]{X.-Y.\ Zeng (曾翔宇)%
       $^{\href{https://orcid.org/0009-0006-1963-7388}{\textcolor[rgb]{0.00,1.00,0.00}{\sf ID}},}$}

\author[NJU,INP]{Y.-Y.\ Xiao (肖宇洋)%
       $^{\href{https://orcid.org/0009-0006-1963-7388}{\textcolor[rgb]{0.00,1.00,0.00}{\sf ID}},}$}

\author[UHe]{Z.-N.\ Xu (徐珍妮)%
    $^{\href{https://orcid.org/0000-0002-9104-9680}{\textcolor[rgb]{0.00,1,0.00}{\sf ID}},}$}

\author[NJU,INP]{C.\ D.\ Roberts%
       $^{\href{https://orcid.org/0000-0002-2937-1361}{\textcolor[rgb]{0.00,1.00,0.00}{\sf ID}},}$}

\author[UHe]{J.\ Rodr\'{\i}guez-Quintero%
       $^{\href{https://orcid.org/0000-0002-1651-5717}{\textcolor[rgb]{0.00,1,0.00}{\sf ID}},}$}

\address[NJU]{
School of Physics, \href{https://ror.org/01rxvg760}{Nanjing University}, Nanjing, Jiangsu 210093, China}
\address[INP]{
Institute for Nonperturbative Physics, \href{https://ror.org/01rxvg760}{Nanjing University}, Nanjing, Jiangsu 210093, China}

\address[UHe]{Department of Integrated Sciences and Center for Advanced Studies in Physics, Mathematics and Computation, \href{https://ror.org/03a1kt624}{University of Huelva}, E-21071 Huelva, Spain
\\[1ex]
\href{mailto:zhenni.xu@dci.uhu.es}{zhenni.xu@dci.uhu.es} (ZNX);
\href{mailto:cdroberts@nju.edu.cn}{cdroberts@nju.edu.cn} (CDR)
\\[1ex]
Date: 2026 April 07\\[-6ex]
}

\begin{abstract}
Charmonia are often supposed to provide simple hydrogen-like ``atomic'' systems that can be used to obtain insights into heavier-quark QCD.
We use continuum Schwinger function methods to analyse this hypothesis in connection with ground-state scalar and pseudoscalar charmonia and find that a more complex picture of these states may be necessary.
For instance, considering orbital angular momentum, the $\chi_{c0}$ is not a simple $P$-wave system; similarly, the $\eta_c$ wave function contains more than merely $S$-wave contributions.
The distribution amplitudes (DAs) and distribution functions (DFs) of these mesons are also nontrivial.
For instance, the $\chi_{c0}$ DA is not positive definite: owing to QCD symmetries, it possesses domains of balanced negative and positive support.
This feature is also expressed in the $\chi_{c0}$ DF, but differences between $\chi_{c0}$ and $\eta_c$ DFs diminish under scale evolution.
Notably, the light-front momentum fraction carried by glue is the same in both states: it is 10\% less than the in-pion glue momentum fraction.
Whilst experimental confirmation of the predictions herein is unlikely, our results should serve as benchmarks for complementary theory attempts to understand local and global structural features of heavier-quark hadrons.
\end{abstract}


\end{frontmatter}
\end{CJK*}

\section{Introduction}
The charm quark, $c$, is special.
With an effective current mass $m_c \approx 1.3\,$GeV \cite{ParticleDataGroup:2024cfk}, just 40\% greater than $m_p$, the proton mass, the $c$ quark is neither light nor truly heavy.
Consequently, $c \bar c$ mesons are states within which mass generated solely by Higgs boson couplings into QCD is fairly well balanced with
that arising from constructive interference between the Higgs-generated current-mass and dynamical effects, \emph{i.e}., emergent hadron mass (EHM), and/or EHM alone; see, \emph{e.g}., Ref.\,\cite[Table~1, Fig.\,18]{Ding:2022ows}.

There are various theoretical approaches to the treatment of $c \bar c$ mesons.
Quark (potential) models have long been used \cite[Ch.\,15]{ParticleDataGroup:2024cfk}.
In these frameworks, one often views such systems from the meson rest frame and therein defines total quark + antiquark spin $\mathpzc s$, and orbital angular momentum (OAM), $\mathpzc l$.
Then, there are strict identities for system parity, $P$, and charge-conjugation parity, $C$:
$P=(-1)^{\mathpzc l + 1}$; $C=(-1)^{\mathpzc l + \mathpzc s}$.
Consequently, low-lying pseudoscalar and vector mesons are considered to be purely $S$-wave systems, \emph{viz}.\ $^1S_0$, $^3 S_1$, respectively; and scalar mesons are pictured as purely $P$-wave states, $^3P_0$.

On the basis of Poincar\'e covariance, such assignments are known to be flawed for mesons built from the lighter quarks, $u$, $d$, $s$; see, \emph{e.g}., Refs.\,\cite{Hilger:2015ora, Xiao:2025cqz}.
On the other hand, it is common to suppose that the nonrelativistic interpretative scheme is a fair, even good, representation for $c \bar c$ states.  Nonrelativistic QCD (NRQCD) \cite{Brambilla:1999xf} may be seen as one formalisation of this perspective.

Herein, we address properties of $c \bar c$ mesons (charmonia) using continuum Schwinger function methods (CSMs).
CSMs deliver a fully Poincar\'e-covariant approach that also preserves the other symmetries that are important to the discussion of mesons.
Our study extends Refs.\,\cite{Hilger:2014nma, Fischer:2014cfa, Yin:2021uom, Raya:2016yuj, Chen:2016bpj} via the following novel contributions:
it delivers the Bethe-Salpeter wave functions (BSWFs) of both ground-state scalar and pseudoscalar quarkonia, \emph{i.e}., the lowest mass parity partners in the charmonia sector;
therewith examines their rest-frame OAM content;
and therefrom calculates and contrasts their leading-twist quasiparticle distribution amplitudes (DAs) and distribution functions (DFs -- valence, glue, and sea).

Initially, we deliver predictions at the hadron scale, $\zeta_{\cal H}<m_p$.
At $\zeta_{\cal H}$, all properties of a given hadron are carried by its quasiparticle valence degrees of freedom (dof).
The existence of a hadron scale is guaranteed by the theory of QCD effective charges \cite{Grunberg:1980ja, Grunberg:1982fw}, \cite[Sec.\,4.3]{Deur:2023dzc}.
We subsequently evolve DF results to scales $\zeta > \zeta_{\cal H}$ using the all-orders (AO) scheme \cite{Yin:2023dbw}.
This nonperturbative extension of the DGLAP evolution approach \cite{Dokshitzer:1977sg, Gribov:1971zn, Lipatov:1974qm, Altarelli:1977zs} has proven valuable in many applications; see, \textit{e.g}., Refs.\,\cite{Han:2020vjp, Lu:2022cjx, Wang:2023bmk, Xu:2023bwv, Lu:2023yna, Xu:2024nzp, Yao:2024ixu, Castro:2025rpx, Xing:2025eip, Cheng:2026nud}.

\section{$c\bar c$ mesons via the Bethe-Salpeter equation}
\label{SecBSWF}
Light scalar mesons are  not well described as bound-states built primarily from a quasiparticle-quark + quasiparticle-antiquark \cite{Pelaez:2015qba}: for these light systems, resonant (light-meson + light-meson scattering) terms couple directly into the Bethe-Salpeter kernel; hence, are significant \cite{Holl:2005st, Eichmann:2015cra, Xu:2022kng}.
Such contributions are much suppressed in ground-state $c \bar c$ systems because they do not share valence dof in common with light mesons.
This conclusion is supported, \emph{e.g}., by the fact that, in quasiparticle-quark + quasiparticle-antiquark Bethe-Salpeter equation treatments of $c \bar c$ mesons, one readily reproduces the correct masses and level ordering of $\eta_c$, $J/\psi$, $\chi_{c0}$ \cite{Hilger:2014nma, Fischer:2014cfa, Yin:2021uom}.

The Poincar\'e-covariant BSWF for the $\chi_{c0}$, a $J^{PC}=0^{++}$, $c\bar c$ state, can be written in the following form:
{\allowdisplaybreaks
\begin{subequations}
\label{BSWF}
\begin{align}
{\mathpzc X}_{\mathsf 0}& (k;Q)  =
S_c(k_\eta) \Gamma_0(k;Q) S_c(k_{\bar\eta})\\
& = S_c(k_\eta)  {\mathbb I} \big[
 {E}_{\mathsf 0}(k;Q) + i k\cdot Q \gamma\cdot Q {F}_{\mathsf 0}(k;Q) \nonumber \\
&
\qquad + \! i \gamma\cdot k \,  {G}_{\mathsf 0}(k;Q) + i \sigma_{\mu\nu} k_\mu Q_\nu {H}_{\mathsf 0}(k;Q) \big]S_c(k_{\bar\eta})
\label{Eq1b}\\
& =: {\mathbb I} \big[
 {\cal E}_{\mathsf 0}(k;Q) + i k\cdot Q\gamma\cdot Q {\cal F}_{\mathsf 0}(k;Q) \nonumber \\
&
\qquad +\!  i \gamma\cdot k \, {\cal G}_{\mathsf 0}(k;Q) + i \sigma_{\mu\nu} k_\mu Q_\nu {\cal H}_{\mathsf 0}(k;Q) \big]\,, \label{XOAM} \\
&
 =: {\mathpzc g}^1_{\mathsf 0} {\mathpzc E}_{\mathsf 0}(k;Q) + {\mathpzc g}^2_{\mathsf 0} {\mathpzc F}_{\mathsf 0}(k;Q) \nonumber \\%
&
\qquad + {\mathpzc g}^3_{\mathsf 0} {\mathpzc G}_{\mathsf 0}(k;Q) + {\mathpzc g}^4_{\mathsf 0} {\mathpzc H}_{\mathsf 0}(k;Q)\,, \label{XOAM2} \\
&
=: {\mathpzc X}_{\mathsf 0}^{1}(k;Q) + {\mathpzc X}_{\mathsf 0}^{2}(k;Q) + {\mathpzc X}_{\mathsf 0}^{3}(k;Q) + {\mathpzc X}_{\mathsf 0}^{4}(k;Q) \,,\label{XOAM3}
\end{align}
\end{subequations}
where
$S_{c}$ is the $2$-point Schwinger function (propagator) for the valence $c$, $\bar c$ constituents of the meson;
$\Gamma_0(k;Q)$ is the $\chi_{c0}$ BS amplitude (amputated BSWF);
$Q$ is the meson total momentum, $Q^2 = -m_{\mathsf 0}^2$, $m_{\mathsf 0}$ is the $\chi_{c0}$ mass;
$k$ is the relative momentum between the valence $c$ and $\bar c$, whose properties specify the character of the system;
$k_\eta = k+\eta Q$, $k_{\bar\eta} = k-(1-\eta) Q$, $0\leq \eta \leq 1$;
and ${\mathpzc g}_{\mathsf 0}^{i=1,2,3,4}$ are Dirac matrices, defined implicitly by comparing Eqs.\,\eqref{XOAM}, \eqref{XOAM2}.
It is useful to choose $\eta=1/2$; then, $E_{\mathsf 0}, F_{\mathsf 0}, G_{\mathsf 0}, H_{\mathsf 0}$ in Eq.\,\eqref{Eq1b} are even functions of $k\cdot Q$.
The analogous BSWF for the pseudoscalar $\eta_c$-meson can be inferred from Ref.\,\cite[Eq.\,(1)]{Xiao:2025cqz}.
}

Working with a rest-frame projection of the BSWF in Eq.\,\eqref{BSWF}, one finds \cite{Hilger:2015ora} that ${\cal E}_{\mathsf 0}, {\cal F}_{\mathsf 0}$ correspond to $S$-wave, whereas ${\cal G}_{\mathsf 0}, {\cal H}_{\mathsf 0}$ are $P$-wave components.
It follows that if a quark model picture is valid, then ${\cal G}_{\mathsf 0}$ and/or ${\cal H}_{\mathsf 0}$ should dominate the wave function.

The BSWF can be obtained by solving a homogeneous Bethe-Salpeter equation, which may be written as follows:
\begin{equation}
\Gamma_{\mathsf 0}(k;Q)
= \lambda(Q^2) \int_{dq} {\mathpzc X}_{\mathsf 0} (q;Q)\, {\cal K}(q,k;Q)\,.
\end{equation}
where $\int_{dq}$ indicates a translationally invariant regularisation of the four-dimensional integral, ${\cal K}$ is the Bethe-Salpeter kernel and $\lambda(P^2)$ is the eigenvalue.
The required meson wave function is obtained at that value of $Q^2$ for which $\lambda(Q^2)=1$ and one reads the mass from this value of $Q^2 = -m_0^2$.

To calculate any observable for comparison with measurement, the canonically normalised BSWF must be used \cite[Sec.\,3]{Nakanishi:1969ph}.
The wave function of a nonrelativistic quantum mechanics bound state, $\Psi(x)$, is a probability amplitude.
Its normalisation is implemented by introducing a multiplicative scaling factor that ensures $1= \int dx\,|\Psi(x)|^2$.
In Poincar\'e-invariant quantum field theory, however, owing to, amongst other things, the loss of particle number conservation, the Poincar\'e-covariant BSWF does not permit interpretation as a probability amplitude; hence, its normalisation is different from that used in quantum mechanics.
Canonical BSWF normalisation is accomplished by rescaling such that \cite[Sec.\,3]{Nakanishi:1969ph}:
{\allowdisplaybreaks
\begin{align}
\label{eq:Nakanishi Normalization}
1 & = \bigg[\frac{d \ln\lambda(Q^2)}{d Q^2} \times
\nonumber \\
& \times \textrm{tr}_\textrm{CD}\int_{dk}
\bar{\mathpzc X}_{\mathsf 0}(k;-Q)
S_c^{-1}(k_\eta) {\mathpzc X}_{\mathsf 0}(k;Q)S_c^{-1}(k_{\bar \eta})
\bigg]_{Q^2+m_{\mathsf 0}^2=0}\,,
%
%
%
\end{align}
where the trace is over colour and spinor indices,
and the charge-conjugated BSWF is $\bar{\mathpzc X}_{\mathsf 0}(k; -Q) = C^\dagger {\mathpzc X}_{\mathsf 0}^{\rm T}(-k; -Q) C$, with $C=\gamma_2 \gamma_4$ and $(\cdot)^T$ indicating matrix transpose \cite[Eq.\,(27)]{Maris:1997tm}.
%
(See Ref.\,\cite[Fig.\,3]{Wang:2018kto} for the baryon analogue.)
}

One observable that is commonly used to characterise a meson is its leptonic decay constant.  In the present case,
\begin{equation}
f_{\mathsf 0} Q_\mu =
{\rm tr}_{\textrm{CD}}\int_{dk} \gamma_\mu {\mathpzc X}_{\mathsf 0}(k;Q)\,.
\label{f0zeroO}
\end{equation}
As may be inferred using the vector Ward-Green-Takahashi identity,
\begin{equation}
f_{\mathsf 0} \equiv 0
\label{f0zero}
\end{equation}
for any $0^{++}$ bound state built from mass-degenerate valence dof \cite{Maris:2000ig}.  Recovering this outcome is a check on whether any given approach is symmetry-preserving.

The $c$ quark propagator in Eq.\,\eqref{BSWF} has the following form:
\begin{subequations}
\label{Sprop}
\begin{align}
    S_{c}(q)&  = 1/[i \gamma\cdot q A_{c}(q^2)+ B_{c}(q^2) ]\,,\\
    & =: Z_{c}(q^2)/[i \gamma\cdot q + M_{c}(q^2) ]\,. \label{Scprop2}
\end{align}
\end{subequations}
The function $A_c(q^2)$, linked to the vector part of the dressed-quark self-energy, becomes uniformly closer to unity as the quark current mass is increased; but in a sign that the $c$ quark is not truly heavy, deviations from unity are apparent in $A_{c}(q^2)$ over a material $q^2$ domain.
The scalar part of the self-energy, $B_{c}(q^2)$ is also much ``flatter'' than the analogous light-quark function, \emph{i.e}., it is more weakly dependent on $q^2$; additionally,
$B_{c}(q^2)$ is uniformly greater in magnitude.
Looking at Eq.\,\eqref{Scprop2}, $M_c(q^2)$  is the $c$ dressed-mass function, which is renormalisation group invariant (RGI) in QCD \cite{Politzer:1976tv}.
Reviewing \cite[Fig.\,2.5]{Roberts:2021nhw}, one perceives the slow but persistent running of $M_c(q^2)$.

Using CSMs, the solution of a given meson bound state problem is found by considering a set of coupled gap and Bethe-Salpeter equations \cite{Roberts:1994dr, Roberts:2012sv}.
Feedup within this infinite tower of Dyson-Schwinger equations (DSEs \cite{Roberts:1994dr}) makes it necessary to introduce an approximation scheme so that physics results can be obtained from a finite subsystem of equations.
Given the size of the $m_c$, it is reasonable to use the leading-order approximation in the scheme introduced in Refs.\,\cite{Munczek:1994zz, Bender:1996bb}, \emph{viz}.\ rainbow-ladder (RL) truncation.
This is because, with growing current-mass, DSE contributions from nonplanar diagrams and vertex corrections are increasingly suppressed -- see, \emph{e.g}., Ref.\,\cite{Bhagwat:2004hn}; and for the $c$ quark, such modifications can largely be accommodated in a sensibly formulated RL truncation.

In realistic implementations of RL truncation, which can be traced from Ref.\,\cite{Maris:1997tm}, the principal element is the effective charge.
Studies of QCD's gauge sector have delivered the following efficacious form for the product of effective charge and gluon $2$-point function (propagator) \cite{Qin:2011dd, Binosi:2014aea, Binosi:2016wcx, Yao:2024ixu}:
\begin{align}
\label{defcalG}
 {\mathpzc d}(y) & =
 \frac{2\pi}{\omega^4} D e^{-y/\omega^2} + \frac{\pi \gamma_m \mathcal{F}(y)}{\tfrac{1}{2}\ln\big[ \tau+(1+y/\Lambda_{\rm QCD}^2)^2 \big]}\,,
\end{align}
where \cite{Qin:2018dqp, Yao:2021pyf}:
$\gamma_m=12/23$, $\Lambda_{\rm QCD} = 0.36\,$GeV, $\tau={\rm e}^2-1$, and ${\cal F}(y) = \{1 - \exp(-y/\Lambda_{\mathpzc I}^2)\}/y$, $\Lambda_{\mathpzc I}=1\,$GeV.
The kernel of the gap equation for $S_{c}(p)$ can then be written as $(l=(p-q), y=l^2)$ \cite{Maris:1997tm, Binosi:2016wcx}:
\begin{equation}
\label{DSEkernel}
{\mathpzc d}(y) T_{\mu\nu}(l) [i\gamma_\mu\frac{\lambda^{a}}{2} ]_{tr} [i\gamma_\nu\frac{\lambda^{a}}{2} ]_{su}\,,
\end{equation}
$l^2 T_{\mu\nu}(l) = l^2 \delta_{\mu\nu} - l_\mu l_\nu$.
We use Landau gauge because, amongst other strengths, it is a fixed point of the QCD renormalisation group.
In solving all DSEs involved in the problem, we use a mass-independent momentum-subtraction renormalisation scheme \cite{Chang:2008ec}, with renormalisation scale $\zeta=\zeta_{19}:=19\,$GeV.  At $\zeta_{19}$, one finds that the flavour-independent quark wave function renormalisation constant $Z_2\approx 1$, a feature that leads to useful simplifications.
Evolution to other renormalisation scales is straightforward.

Widespread use has revealed that, when the value of the product $\varsigma_Q^3=\omega D$ is kept fixed, many ground-state hadron observables are practically insensitive to changes $\omega \to (1\pm 0.1)\omega$ \cite{Qin:2018dqp}.
So, considering Refs.\,\cite{Qin:2018dqp, Yao:2021pyf}, which, respectively, provided successful descriptions of heavy quark baryons and $B_c \to \eta_c, J/\psi$ transitions, we use
\begin{equation}
\omega = 0.8\,{\rm GeV}\,,
\quad \varsigma_Q = 0.6\,{\rm GeV}\,.
\end{equation}
Then with a RGI $c$ current mass $\hat m_c = 1.61\,$GeV and solving the necessary gap and Bethe-Salpeter equations using standard algorithms \cite{Maris:1997tm, Maris:2005tt, Krassnigg:2009gd}, one obtains the following results:
$M_c(\zeta_2^2) = 1.27\,$GeV, $\zeta_2=2\,$GeV;
a Euclidean constituent mass $M_c^E = 1.33\,$GeV;
\begin{equation}
\begin{array}{l|lcc}
& {\mathsf M}  & m_{\mathsf M} &  f_{\mathsf M} \\\hline
{\rm herein}\
    & \chi_{c0} & 3.33 & 0 \\
    & \eta_c & 2.98 & 0.28 \\\hline
\mbox{\cite[PDG]{ParticleDataGroup:2024cfk}}
    & \chi_{c0} & 3.41 & 0 \\
    & \eta_c & 2.98 & 0.24 \\\hline
\end{array}\;.
\end{equation}
(The value of $f_{\eta_c}$ was extracted from information on the decay width for $\eta_c \to \gamma\gamma$ decays \cite[PDG]{ParticleDataGroup:2024cfk}; see also Ref.\,\cite{Raya:2016yuj}.)
In addition, one obtains the $\chi_{c0}$, $\eta_c$ BSWFs.

\section{OAM in the rest frame}
\label{Sec3}
Consider the following set of projection operators:
\begin{subequations}
\label{projectors}
    \begin{align}
    {\mathcal P}^{\mathsf 0}_{1={\cal E} = S_{\rm wave}} & = \tfrac{1}{4}{\mathbb I}  \,,\\
    {\mathcal P}^{\mathsf 0}_{2={\cal F} = S_{\rm wave}} & = \tfrac{-i}{4} \tfrac{1}{Q^2}\tfrac{1}{k\cdot Q} \gamma\cdot Q \,, \\
    {\mathcal P}^{\mathsf 0}_{3={\cal G} = P_{wave}} & = \tfrac{-i}{4}\tfrac{1}{k^2} \gamma\cdot k \,,\\
    {\mathcal P}^{\mathsf 0}_{4={\cal H} = P_{\rm wave}} & =\tfrac{i}{4}\tfrac{1}{(k\cdot Q)^2-k^2 Q^2} \sigma_{\mu\nu} Q_\mu k_\nu \,.
    \end{align}
\end{subequations}
Applied to Eq.\,\eqref{XOAM}, they yield the separate wave function components whose rest-frame OAM association is specified in the subscript.
Working with these operators, one has (no sum on $i$):
\begin{equation}
{\mathpzc X}_{\mathsf 0}^i (k;Q) = {\mathpzc g}_{\mathsf 0}^i {\rm tr}_{\rm D}{\mathcal P}^{\mathsf 0}_{i}{\mathpzc X}_{\mathsf 0}(k;Q)\,.
\end{equation}
Now, construct the following matrix:
\begin{align}
    & {\mathpzc L}^{ij}
    = - \left[ \frac{d \ln\lambda(Q^2)}{d Q^2} \right. \nonumber \\
&    \left. \times \, \textrm{tr}_\textrm{CD}\int_{dk}
\bar{\mathpzc X}_{\mathsf 0}^i(k;-Q)
S_c^{-1}(k_\eta) {\mathpzc X}_{\mathsf 0}^j(k;Q)S_c^{-1}(k_{\bar \eta})
\right]_{P^2+m_{\mathsf 0}^2=0}\,,
\label{DefLIJ}
\end{align}
wherewith Eq.\,\eqref{eq:Nakanishi Normalization} guarantees
    $1 = \sum_{i,j}^4 {\mathpzc L}^{ij}\,.$
This matrix provides a measure of the rest-frame OAM pairing contributions to a meson's canonical normalisation in a form that is somewhat like the bound-state wave function normalisation condition in quantum mechanics.

The $\eta_c$ analogues are readily inferred from Ref.\,\cite[Sect.\,3]{Xiao:2025cqz}.

\begin{figure}[t]
\vspace*{1ex}

\hspace*{-1ex}\begin{tabular}{l}
\large{\textsf{A}$\quad \chi_{c0}$}\\[-2.ex]
%
\centerline{\includegraphics[clip, width=0.3\textwidth]{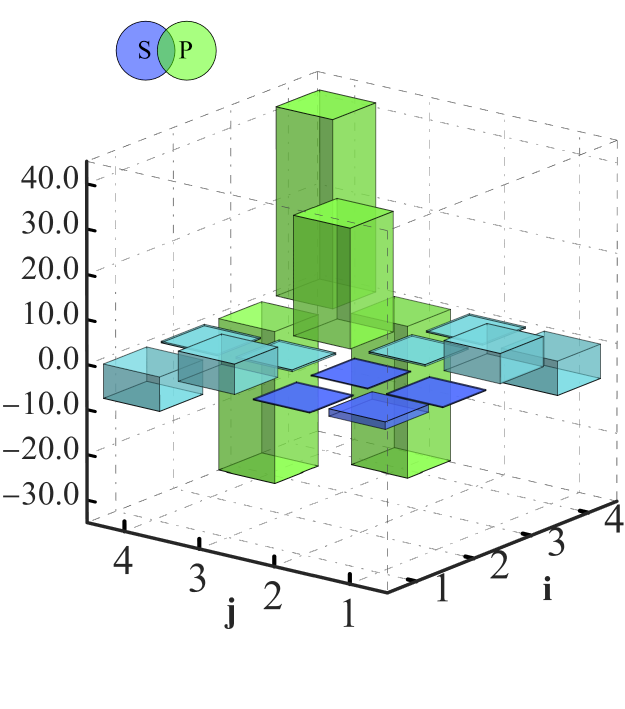}}
\end{tabular}
\hspace*{-1ex}\begin{tabular}{l}
\large{\textsf{B}$\quad \eta_c$} \\[0.ex]
\centerline{\includegraphics[clip, width=0.3\textwidth]{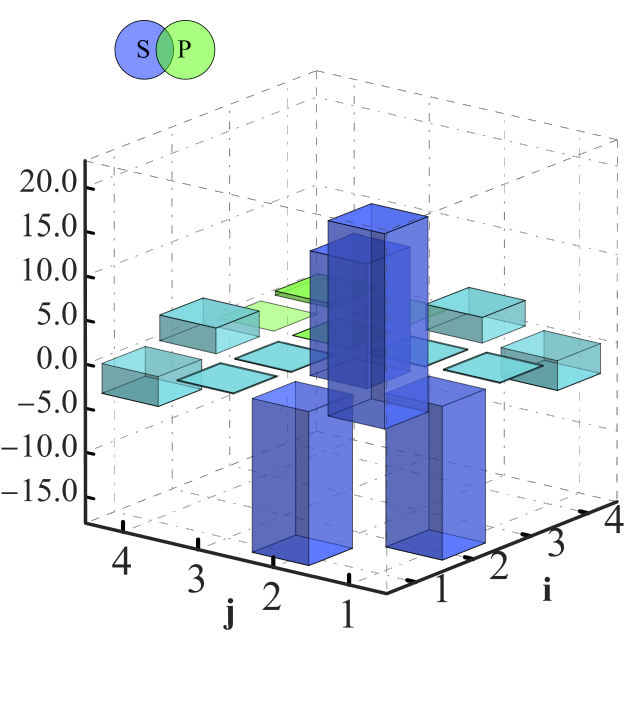}}
\end{tabular}
\vspace*{-4ex}

\caption{\label{F1}
Rest-frame OAM decompositions of $\chi_{c0}$, $\eta_c$ BSWFs as defined via Eqs.\,\eqref{projectors}\,-\,\eqref{DefLIJ}.
There are both positive (above plane) and negative (below plane) contributions, the total sum of which is unity in each case.
}
\end{figure}

The calculated $\chi_{c0}$ OAM decomposition obtained using Eq.\,\eqref{DefLIJ} is presented in Fig.\,\ref{F1}A.
Evidently, the wave function OAM content is highly nontrivial; so even for this $c\bar c$ system, a simple quark model wave function is a poor approximation.
The $P\otimes P$-wave contributions are very large; but, as apparent in Table~\ref{OAMD}, they cancel amongst themselves;
the individual magnitudes of the $S \otimes S$-wave pieces are much smaller, but they sum to a greater amount;
and the largest single net contribution to the normalisation is generated by $S \otimes P$ interference.
A similar picture is seen for lighter scalar mesons \cite[Fig.\,7]{Hilger:2015ora}.

\begin{table}[t]
\caption{ \label{OAMD}
Contributions (measured in \%) from various in-meson rest-frame partial wave overlaps to the meson canonical normalisation as measured by the matrix in Eq.\,\eqref{DefLIJ}.  The entries quantify the images in Fig.\,\ref{F1}.
The final row lists results calculated for a (fictitious) pseudoscalar state constituted from two valence dof with current masses which match that of the $s$ quark \cite{Xiao:2025cqz}.
}
\begin{tabular*}
{\hsize}
{
l@{\extracolsep{0ptplus1fil}}
|c@{\extracolsep{0ptplus1fil}}
c@{\extracolsep{0ptplus1fil}}
c@{\extracolsep{0ptplus1fil}}}\hline\hline
\centering
Eq.\,\eqref{DefLIJ} $\ $ & $S \otimes S\ $ & $S \otimes P\ $ & $P \otimes P\ $ \\\hline
$\chi_{c0}\ $ & $142\ $ & $-172\ $ & $130 \ $ \\
$\eta_c\ $ & $141\ $ & $-92\ $ & $51 \ $ \\
$\pi_{s\bar s}\ $ & $109\ $ & $-22\ $ & $13 \ $ \\
\hline\hline
\end{tabular*}
\end{table}

The analogous $\eta_c$ OAM decomposition is drawn in Fig.\,\ref{F1}B.
Again, the wave function OAM content is nontrivial; so a simple quark model wave function is not a good approximation in this case, either.
The $S \otimes S$-wave pieces are large, but roughly a factor of $2$ smaller in magnitude than the $\chi_{c0}$ $P\otimes P$-wave terms;
the $P \otimes P$-wave parts in the $\eta_c$ are significantly smaller than $S \otimes S$ and add to a smaller amount;
and here, too, there is a large contribution to the $\eta_c$ normalisation from $S \otimes P$ interference.

For reference, the final row of Table~\ref{OAMD} lists results for a (fictitious) pseudoscalar state built from two valence dof with current masses which match that of the $s$ quark \cite{Xiao:2025cqz}.
They reveal both the impact of increasing the current mass and changing the $J$ value.

\section{Parton distribution amplitudes}
\label{secDA}
The $\eta_c$ leading-twist valence dof DA was first calculated in Ref.\,\cite{Ding:2015rkn}.  Herein, we revisit that analysis and extend it with calculation of the analogous $\chi_{c0}$ DA.  The latter will be our vehicle for presenting the computational method.
Reviewing Ref.\,\cite{Ding:2015rkn}, it is plain that one should expect charmonia DAs to appear much suppressed on the endpoint domains, $x\simeq 0,1$, when compared with the QCD asymptotic DA \cite{Lepage:1980fj}: $\varphi_{\rm as}=6 x(1-x)$.

Following Ref.\,\cite{Li:2016mah}, the $\chi_{c0}$ DA can be obtained by projection of the meson's BSWF onto the light-front, \emph{viz}.\
\begin{align}
\tilde f_{\mathsf 0} \varphi_{\mathsf 0}(x) & =  N_{c} {\rm tr}_{\rm D}
\int_{d k} \delta\left(n \cdot k_{1/2}-x n \cdot Q\right) \gamma \cdot n {\mathpzc X}_{\mathsf 0}(k ; Q)\,,
\end{align}
where $n$ is a lightlike $4$-vector, $n^2=0$,
and $n\cdot Q=-m_{\mathsf 0}$ in the meson rest frame.
Owing to Eq.\,\eqref{f0zero}, a symmetry constraint, one cannot here factor out a nonzero, measurable decay constant to set the mass-scale.
Therefore, we have introduced the quantity $\tilde f_{\mathsf 0}$, which has mass dimension unity and, in the context of the following Mellin moments:
\begin{equation}
\langle x^m \rangle_{\varphi_{\mathsf 0}}
= \int_0^1 dx\, x^m \varphi_{\mathsf 0}(x)\,,
\label{Mellin}
\end{equation}
takes a value that ensures $\langle x \rangle_{\varphi_{\mathsf 0}} = 1$.  It should also be noted that,  owing to charge-conjugation symmetry \cite{Li:2016mah} and also underlying Eq.\,\eqref{f0zero}:
\begin{equation}
0^{++}\!: \quad \varphi_{\mathsf 0}(1-x) = - \varphi_{\mathsf 0}(x)\,.
\label{Eqantisymm}
\end{equation}

Expressed in terms of the $\chi_{c0}$ BSWF, the Mellin moments in Eq.\,\eqref{Mellin} translate into the following expression:
\begin{align}
\tilde f_{\mathsf 0} \langle x^m \rangle_{\varphi_{\mathsf 0}} (n\cdot Q)^{m+1}  & =
\int_{d k} (n\cdot k_{1/2})^m \gamma \cdot n {\mathpzc X}_{\mathsf 0}(k ; Q)\,.
\label{DirectMM}
\end{align}
Since $\hat m_c > m_p$, one can reliably obtain the first five nontrivial Mellin moments by brute force, \emph{i.e}., unrefined, direct calculation using interpolations of the numerical results for the scalar functions in the BSWF.
(For lighter quarks, more sophisticated methods are required \cite{Li:2016mah, Xu:2025cyj}.)
The $\chi_{c0}$ moments thus obtained are collected in Table~\ref{TabMellin}A.

\begin{table}[t]
\caption{ \label{TabMellin}
{\sf Panel A}.
Mellin moments of the $\chi_{c0}$ DA as obtained directly from the BSWF, using Eq.\,\eqref{DirectMM}, and compared with those produced by the reconstruction form, Eq.\,\eqref{fitalpha}.
$\tilde f_0 = 0.032\,$GeV.
{\sf Panel B}. Analogous moments for the $\eta_c$.
}
\begin{tabular*}
{\hsize}
{
l@{\extracolsep{0ptplus1fil}}
|c@{\extracolsep{0ptplus1fil}}
c@{\extracolsep{0ptplus1fil}}
c@{\extracolsep{0ptplus1fil}}
c@{\extracolsep{0ptplus1fil}}
c@{\extracolsep{0ptplus1fil}}
c@{\extracolsep{0ptplus1fil}}}\hline\hline
\centering
{\sf A}. $\quad m\ $ & $0\ $ & $1\ $ & $2\ $ & $3\ $ & $4\ $ & $5\ $ \\\hline
Eq.\,\eqref{DirectMM} $\ $  & $0\ $ &  $1\ $  & $1\ $  & $0.783\ $  & $0.562\ $  & $0.391\ $\\
Eq.\,\eqref{fitalpha} $\ $ & $0\ $ & $0.984\ $ & $0.984\ $ & $0.774\ $ & $0.564\ $ & $0.399\ $
\\\hline\hline
\end{tabular*}

\medskip

\begin{tabular*}
{\hsize}
{
l@{\extracolsep{0ptplus1fil}}
|c@{\extracolsep{0ptplus1fil}}
c@{\extracolsep{0ptplus1fil}}
c@{\extracolsep{0ptplus1fil}}
c@{\extracolsep{0ptplus1fil}}
c@{\extracolsep{0ptplus1fil}}
c@{\extracolsep{0ptplus1fil}}}\hline\hline
\centering
{\sf B}. $\quad m\ $ & $0\ $ & $1\ $ & $2\ $ & $3\ $ & $4\ $ & $5\ $ \\\hline
Eq.\,\eqref{DirectMMeta} $\ $  & $1\ $ &  $0.5\ $  & $0.285\ $  & $0.177\ $  & $0.116\ $  & $0.080\ $\\
Eq.\,\eqref{fitalpha} $\ $ & $1\ $ & $0.5\ $ & $0.284\ $ & $0.176\ $ & $0.116\ $ & $0.080\ $
\\\hline\hline
\end{tabular*}
\end{table}

\begin{table}[t]
\caption{ \label{TabGegenbauer}
Gegenbauer coefficients for meson DAs.
{\sf Panel A}.
$\chi_{c0}$, Eq.\,\eqref{QCDDA}.  (Recall $\tilde f_0 = 0.032\,$GeV.)
{\sf Panel B}.
$\eta_c$, from the even-Gegenbauer analogue of Eq.\,\eqref{QCDDA}.
}
\begin{tabular*}
{\hsize}
{
l@{\extracolsep{0ptplus1fil}}
|c@{\extracolsep{0ptplus1fil}}
c@{\extracolsep{0ptplus1fil}}
c@{\extracolsep{0ptplus1fil}}
c@{\extracolsep{0ptplus1fil}}
c@{\extracolsep{0ptplus1fil}}
c@{\extracolsep{0ptplus1fil}}
c@{\extracolsep{0ptplus1fil}}
c@{\extracolsep{0ptplus1fil}}}\hline\hline
\centering
A$\ $ &
$a_1\ $ & $a_3\ $ & $a_5\ $ & $a_7\ $ & $a_9\ $ & $a_{11}\ $ & $a_{13}\ $ & $a_{15}\ $ \\
& $3.279\ $ & $-2.921\ $ & $1.802\ $ & $-0.803\ $ & $0.254\ $ & $-0.054\ $ & $0.0068\ $ & $\approx 0\ $
\\\hline\hline
\end{tabular*}

\medskip

\begin{tabular*}
{\hsize}
{
l@{\extracolsep{0ptplus1fil}}
|c@{\extracolsep{0ptplus1fil}}
c@{\extracolsep{0ptplus1fil}}
c@{\extracolsep{0ptplus1fil}}
c@{\extracolsep{0ptplus1fil}}
c@{\extracolsep{0ptplus1fil}}
c@{\extracolsep{0ptplus1fil}}
c@{\extracolsep{0ptplus1fil}}
c@{\extracolsep{0ptplus1fil}}}\hline\hline
\centering
B $\ $ & $a_0\ $ & $a_2\ $ & $a_4\ $ & $a_6\ $ & $a_{8}\ $ & $a_{10}\ $ & $a_{12}\ $ & $a_{14}\ $ \\
 & $1\ $ & $-0.189\ $ & $0.0635\ $ & $-0.0008\ $ & $\approx 0\ $ & $\approx 0\ $ & $\approx 0\ $ & $\approx 0\ $
\\\hline\hline
\end{tabular*}
\end{table}

\begin{figure}[t]
\vspace*{1ex}

\hspace*{-1ex}\begin{tabular}{l}
%
%
\centerline{\includegraphics[clip, width=0.44\textwidth]{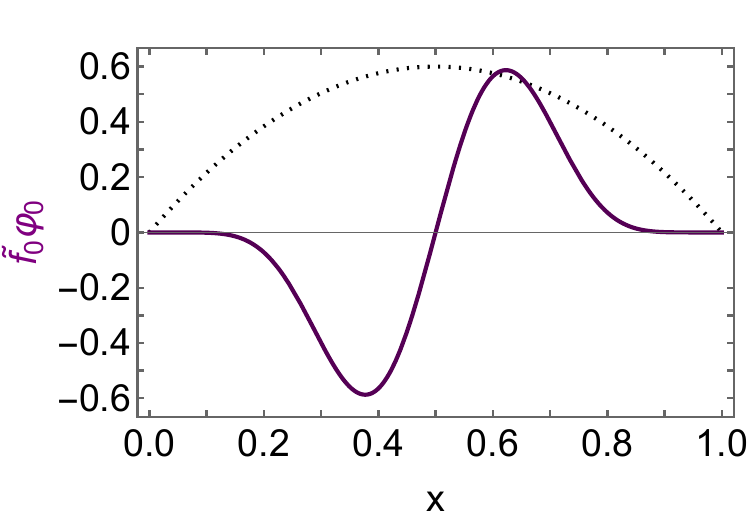}} \\[-37ex]
\large{\textsf{A}$\quad \chi_{c0} \; \mbox{\rm\small DA}$}
\end{tabular}
\vspace*{30.5ex}

\hspace*{-1ex}\begin{tabular}{l}
\centerline{\includegraphics[clip, width=0.44\textwidth]{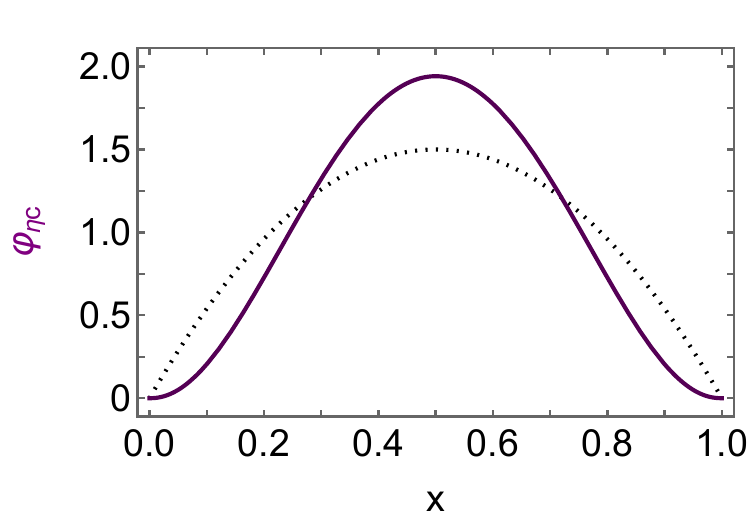}}\\[-37ex]
\large{\textsf{B}$\quad \eta_c \; \mbox{\rm\small DA}$}
\end{tabular}
\vspace*{31ex}
\caption{\label{CompareX0DAs}
Leading-twist valence dof DAs:
{\sf Panel A} -- $\chi_{c0}$ (solid purple) and $(2/5)\varphi_{\rm as}$ (dotted black);
{\sf Panel B} -- $\eta_c$ (solid purple) and $\varphi_{\rm as}$ (dotted black).
}
\end{figure}

We reconstruct the DA from the moments in Table~\ref{TabMellin} by using a two-step approach.
Given that one expects an endpoint suppressed DA, which would be difficult to reconstruct using just a few terms in the standard QCD Gegenbauer expansion, we begin with the following \emph{Ansatz}:
\begin{equation}
\check\varphi_{\mathsf 0}(x) =
c_1 [x(1-x)]^\alpha C_1^\alpha(2x-1)\,,
\label{fitalpha}
\end{equation}
and determine $\{\alpha, c_1\}$ by requiring a least-squares best-fit to the moments in Table~\ref{TabMellin}A.
This procedure yields $\{\alpha = 7.79 , c_1= 354\,076\}$ and the moments in Table~\ref{TabMellin}\,-\,row~2: the mean value of fit/input for the moments is $0.99(2)$, which signals a very good reconstruction.
It follows that the hadron-scale $\chi_{c0}$ light-front wave function (LFWF) must possess a zero in $x$, the location of which is only weakly dependent on the light-front transverse momentum-squared.

To reestablish contact with the QCD expansion, we write
\begin{equation}
\varphi_{\mathsf 0}(x) =
6 x(1-x)  \sum_{j=1}^{j_{\rm max}} a_{2j-1} \, C_{2j-1}^{3/2}(1-2x)\,,
\label{QCDDA}
\end{equation}
where the $C_{2j-1}^{3/2}$ are Gegenbauer polynomials of order $3/2$ and the coefficients,  $a_{2j-1}$, are obtained by projecting $\check\varphi_{\mathsf 0}(x) $ onto this basis.
Using $j_{\rm max}=8$, one obtains a pointwise reliable QCD reconstruction  with the dimensionless coefficients listed in Table~\ref{TabGegenbauer}A.
This DA is drawn in Fig.\,\ref{CompareX0DAs}A.
The domains of balanced negative and positive support are an expression of Eqs.\,\eqref{f0zero}, \eqref{Eqantisymm}; and the anticipated endpoint suppression is clear.

DA moments for the $\eta_c$ can be obtained from:
\begin{align}
f_{\mathsf 5} \langle x^m \rangle_{\varphi_{\mathsf 5}} (n\cdot Q)^{m+1}  & =
\int_{d k} (n\cdot k_{1/2})^m \gamma_5\gamma \cdot n {\mathpzc X}_{\mathsf 5}(k ; Q)\,.
\label{DirectMMeta}
\end{align}
Our calculated results are listed in Table~\ref{TabMellin}B.  For DA reconstruction, again and for the same reasons, we follow a two-step process, beginning with the following \emph{Ansatz} -- the $0^{-+}$ $\eta_c$ DA is even under $x\leftrightarrow (1-x)$:
\begin{equation}
\varphi_{\mathsf 5}(x) =
[x(1-x)]^\alpha
\frac{\sqrt{\pi } \Gamma (\alpha+1)}{2^{2 \alpha + 1}\Gamma \left(\alpha+\frac{3}{2}\right)}\,,
\label{fitalphaH}
\end{equation}
and determining $\alpha$ via a least-squares best-fit to the moments in Table~\ref{TabMellin}B\,-\,row~1.
This yields $\alpha = 2.20 $ and the moments in Table~\ref{TabMellin}B\,-\,row~2: the mean value of fit/input for the moments is $1.00(1)$, signalling an excellent reconstruction.
Projecting now to extract the analogue of Eq.\,\eqref{QCDDA}, which, in this case, involves only even Gegenbauer polynomials, one obtains the coefficients in Table~\ref{TabGegenbauer}B.
The $\eta_c$ DA is drawn in Fig.\,\ref{CompareX0DAs}B: it is contracted with-respect-to $\varphi_{\rm as}$ and the anticipated endpoint suppression is apparent.

\begin{table}[t]
\caption{ \label{DFMellin}
{\sf Panel A}.
Mellin moments of the $\chi_{c0}$ DF as obtained directly from the BSWF, using Eq.\,\eqref{X0DF}, and compared with those produced by the reconstruction form, Eq.\,\eqref{fitbeta}.
{\sf Panel B}. Analogous moments for the $\eta_c$.
\emph{N.B}.\ Baryon number conservation ensures that, in both cases, the $m=0$ moment is unity.
Similarly, at $\zeta_{\cal H}$, momentum conservation entails that the $m=1$ moment is $1/2$.
Our formulation guarantees these outcomes, so these moments are not listed.
}
\begin{tabular*}
{\hsize}
{
l@{\extracolsep{0ptplus1fil}}
|c@{\extracolsep{0ptplus1fil}}
c@{\extracolsep{0ptplus1fil}}
c@{\extracolsep{0ptplus1fil}}
c@{\extracolsep{0ptplus1fil}}
c@{\extracolsep{0ptplus1fil}}}\hline\hline
\centering
{\sf A}. $\quad m\ $ & $2\ $ & $3\ $ & $4\ $ & $5\ $ & $6\ $  \\\hline
Eq.\,\eqref{X0DF} $\ $  & $0.273\ $ &  $0.158\ $  & $0.096\ $  & $0.061\ $  & $0.040\ $ \\
Eq.\,\eqref{fitbeta} $\ $ & $0.272\ $ & $0.159\ $ & $0.097\ $ & $0.061\ $ & $0.040\ $
\\\hline\hline
\end{tabular*}

\medskip

\begin{tabular*}
{\hsize}
{
l@{\extracolsep{0ptplus1fil}}
|c@{\extracolsep{0ptplus1fil}}
c@{\extracolsep{0ptplus1fil}}
c@{\extracolsep{0ptplus1fil}}
c@{\extracolsep{0ptplus1fil}}
c@{\extracolsep{0ptplus1fil}}}\hline\hline
\centering
{\sf B}. $\quad m\ $ & $2\ $ & $3\ $ & $4\ $ & $5\ $ & $6\ $ \\\hline
Eq.\,\eqref{X0DF}$^{{\mathsf 0}\to \mathsf 5}\ $  & $0.266\ $  & $0.149\ $  & $0.0873\ $  & $0.0532\ $  & $0.0336\ $ \\
Eq.\,\eqref{fitbetaH} $\ $ & $0.266\ $ & $0.149\ $ & $0.0875\ $ & $0.0533\ $ & $0.0335\ $
\\\hline\hline
\end{tabular*}
\end{table}

\section{Parton distribution functions: hadron scale}
\label{DFhadron}
As explained and illustrated in Ref.\,\cite{Ding:2019qlr}, valence dof meson DFs can also be calculated via light-front projections of their BSWFs.  Like DAs, one reconstructs the DF from its moments.  In the present case, the Mellin moments of the quasiparticle valence $c$ DF in the $\chi_{c0}$,
${\mathpzc c}^{\mathsf 0}(x;\zeta_{\cal H})={\mathpzc c}^{\mathsf 0}(1-x;\zeta_{\cal H})$, are given by
\begin{subequations}
\label{X0DF}
\begin{align}
\langle x^m \rangle_{\zeta_{\cal H}}^{c_{\mathsf 0}} (n\cdot Q)^{m+1}
& = N_c {\rm tr}\int_{dk} (n\cdot k_{1/2})^m I^{\mathsf 0}(k;Q;\zeta_{\cal H})\,, \\
I^{\mathsf 0}(k;Q;\zeta_{\cal H}) & =
n\cdot\partial_{k_{1/2}}
[ \bar \Gamma_{\mathsf 0}(k_{1/2};-Q) S_c(k_{1/2})] \nonumber \\
& \qquad \times \Gamma_{\mathsf 0}(k_{\overline{1/2}};Q) S_c(k_{\overline{1/2}})\,.
\end{align}
\end{subequations}
Here, we have explicitly reintroduced the hadron scale, $\zeta_{\cal H}$.

\begin{figure}[t]
\vspace*{1ex}

\hspace*{-1ex}\begin{tabular}{l}
%
%
\centerline{\includegraphics[clip, width=0.44\textwidth]{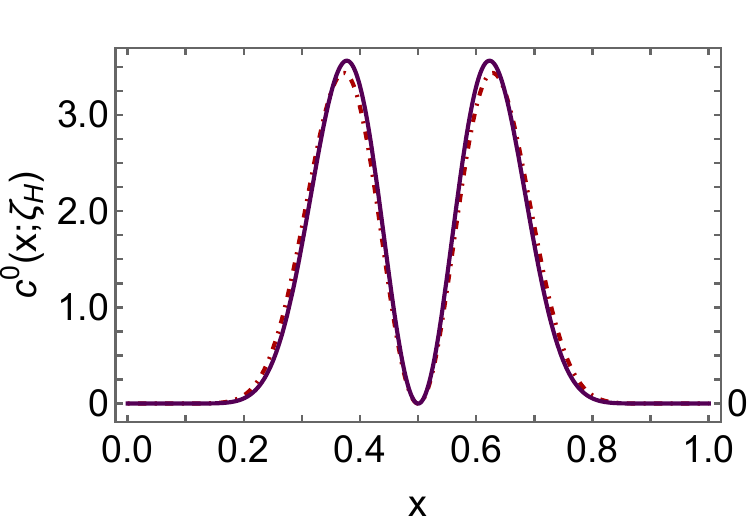}} \\[-37ex]
\large{\textsf{A}$\quad \chi_{c0}\mbox{\small $(\zeta_{\cal H})$} \; \mbox{\rm\small DF}$}
\end{tabular}
\vspace*{30.5ex}

\hspace*{-1ex}\begin{tabular}{l}
\centerline{\includegraphics[clip, width=0.44\textwidth]{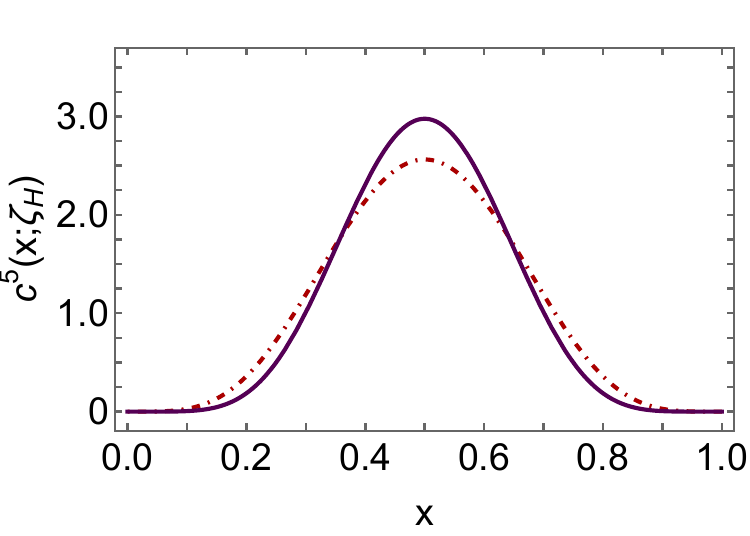}}\\[-37ex]
\large{\textsf{B}$\quad \eta_c\mbox{\small $(\zeta_{\cal H})$} \; \mbox{\rm\small DF}$}
\end{tabular}
\vspace*{31ex}
\caption{\label{CompareX0DFs}
Hadron scale valence dof DFs:
{\sf Panel A} -- $\chi_{c0}$ (solid purple) and zeroth-moment unit-normalised $\varphi_{\mathsf 0}^2$ (dot-dashed red);
{\sf Panel B} -- $\eta_c$ (solid purple) and zeroth-moment unit-normalised $\varphi_{\mathsf 5}^2$ (dot-dashed red)
}
\end{figure}

Using Eq.\,\eqref{X0DF} and brute force, as with the DAs, one can readily calculate the first seven ${\mathpzc c}^{\mathsf 0}(x;\zeta_{\cal H})$ Mellin moments.  The results are listed in Table~\ref{DFMellin}A.

To reconstruct the DF from these moments, we follow the two-step procedure used effectively in Sect.\,\ref{secDA}.
Before proceeding, however, we note that in mesons built from mass-degenerate valence dof, a separable approximation to its LFWF is a good, even very good, approximation when used in connection with integrated quantities and, often, also pointwise \cite{Roberts:2021nhw, Raya:2021zrz, Raya:2024ejx, Xu:2025cyj, Yao:2025xjx}.
It follows that the system's hadron-scale DF should be well approximated by the square of its DA.
So, in step one, we use the following \emph{Ansatz}:
\begin{equation}
\check{\mathpzc c}^{\mathsf 0}(x;\zeta_{\cal H}) =
 [x(1-x)]^\beta [C_1^\beta(2x-1)]^2/{\mathpzc n}_1^{\mathsf 0}\,,
\label{fitbeta}
\end{equation}
where ${\mathpzc n}_1^{\mathsf 0}$ ensures that the zeroth moment is unity, in accordance with baryon number conservation.
Asking for the value of $\beta$ that delivers a least-squares best-fit to the moments in Table~\ref{DFMellin}A\,-\,row~1, one finds $\beta = 14.3$.  This yields the moments in Table~\ref{DFMellin}A\,-\,row~2: the mean value of fit/input for the moments is $1.00(1)$.
Notably, consistent with the separable LFWF hypothesis, $\beta$ is approximately twice the value of $\alpha$ that is associated with the $\chi_{c0}$ DA

Projecting now onto a Gegenbauer-order $5/2$ basis, one finds that a good pointwise representation of the $\chi_{c0}$ valence dof DF is obtained with
\begin{equation}
{\mathpzc c}^{\mathsf 0}(x;\zeta_{\cal H}) =
30 [x (1-x)]^2 \sum_{j=0}^{2 (j_{\rm max}=11)} a_{2j} C_{2j}^{5/2}(2x-1)
\label{DFGegenbauer}
\end{equation}
and the expansion coefficients listed in Table~\ref{TabGegenbauerDF}A.  The $\chi_{c0}$ DF is drawn in Fig.\,\ref{CompareX0DFs}A.  Plainly, the ${\mathpzc c}^{\mathsf 0}(x;\zeta_{\cal H}) \propto \varphi_{\mathsf 0}(x)^2$ hypothesis delivers a very good approximation in this case:
the relative ${\mathpzc L}_1$ difference between the curves is 7.4\%.

\begin{table}[t]
\caption{ \label{TabGegenbauerDF}
Gegenbauer coefficients for meson DFs: Eq.\,\eqref{DFGegenbauer}.
{\sf Panel A} -- $\chi_{c0}$.
{\sf Panel B} -- $\eta_c$.
}
\begin{tabular*}
{\hsize}
{
l@{\extracolsep{0ptplus1fil}}
|c@{\extracolsep{0ptplus1fil}}
c@{\extracolsep{0ptplus1fil}}
c@{\extracolsep{0ptplus1fil}}
c@{\extracolsep{0ptplus1fil}}
c@{\extracolsep{0ptplus1fil}}
c@{\extracolsep{0ptplus1fil}}}\hline\hline
\centering
A$\ $ & $a_0\ $ & $a_2\ $ & $a_4\ $ & $a_6\ $ & $a_8\ $ & $a_{10}\ $ \\
& $1\ $ & $-0.112\ $ & $-0.0316\ $ & $0.0518\ $ & $-0.0381\ $ & $0.0207\ $ \\\hline
& $\mbox{\footnotesize $10$} a_{12}\ $ & $\mbox{\footnotesize $10$}a_{14}\ $
& $\mbox{\footnotesize $10^2$}a_{16}\ $ & $\mbox{\footnotesize $10^2$}a_{18}\ $
& $\mbox{\footnotesize $10^3$}a_{20}\ $ & $a_{22}\ $\\
& $-0.0903\ $ & $0.0320\ $ & $-0.0917\ $ & $0.0209\ $ & $-0.0368\ $ & $\approx 0\ $
\\\hline\hline
\end{tabular*}

\medskip

\begin{tabular*}
{\hsize}
{
l@{\extracolsep{0ptplus1fil}}
|c@{\extracolsep{0ptplus1fil}}
c@{\extracolsep{0ptplus1fil}}
c@{\extracolsep{0ptplus1fil}}
c@{\extracolsep{0ptplus1fil}}
c@{\extracolsep{0ptplus1fil}}
c@{\extracolsep{0ptplus1fil}}
c@{\extracolsep{0ptplus1fil}}
c@{\extracolsep{0ptplus1fil}}}\hline\hline
\centering
B$\ $ & $a_0\ $ & $a_2\ $ & $a_4\ $ & $\mbox{\footnotesize $10$} a_6\ $ & $\mbox{\footnotesize $10^2$} a_8\ $ & $a_{10}\ $ \\
& $1\ $ & $-0.164\ $ & $0.0326\ $ & $-0.0485\ $ & $0.0393\ $ & $\approx 0\ $
%
%
\\\hline\hline
\end{tabular*}
\end{table}

Repeating the steps now for the $\eta_c$, beginning with the \emph{Ansatz}
\begin{equation}
\check{\mathpzc c}^{\mathsf 5}(x;\zeta_{\cal H}) =
 [x(1-x)]^\beta {\mathpzc n}_1^{\mathsf 5}\,,
\label{fitbetaH}
\end{equation}
one finds $\beta = 6.20$, which yields the moments in Table~\ref{DFMellin}B\,-\,row~2: the mean value of fit/input for the moments is $1.000(2)$.
In this case, $\beta$ is approximately thrice the value of the $\alpha$ associated with the $\eta_{c}$ DA.
Nevertheless, as apparent in Fig.\,\ref{CompareX0DFs}B, the separable LFWF hypothesis is a fair pointwise approximation in this case, too: the relative ${\mathpzc L}_1$ difference between the curves is 14\%.
Projecting onto a Gegenbauer-order $5/2$ basis, one finds that a good pointwise representation of the $\eta_c$ valence dof DF is obtained with the expansion coefficients listed in Table~\ref{TabGegenbauerDF}B.

\begin{figure}[t]
\vspace*{1ex}

\hspace*{-1ex}\begin{tabular}{l}
%
%
\centerline{\includegraphics[clip, width=0.44\textwidth]{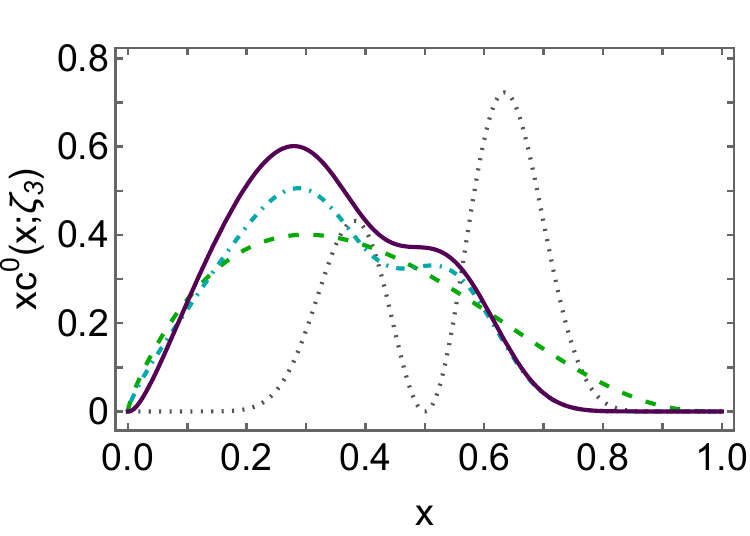}} \\[-37ex]
\large{\textsf{A}$\quad \chi_{c0}\mbox{\small $(\zeta_3)$} \; \mbox{\rm\small DF}$}
\end{tabular}
\vspace*{30.5ex}

\hspace*{-1ex}\begin{tabular}{l}
\centerline{\includegraphics[clip, width=0.44\textwidth]{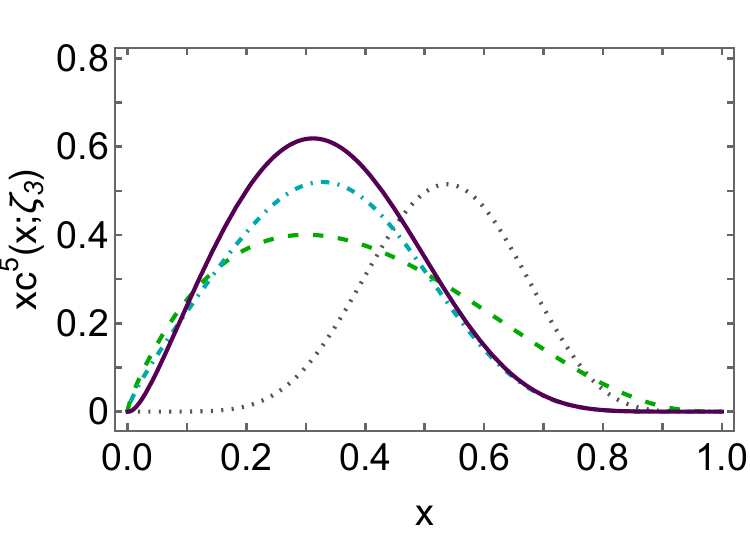}}\\[-37ex]
\large{\textsf{B}$\quad \eta_c\mbox{\small $(\zeta_3)$} \; \mbox{\rm\small DF}$}
\end{tabular}
\vspace*{31ex}
\caption{\label{DFsevolved}
Valence dof DFs in Fig.\,\ref{CompareX0DFs} evolved $\zeta_{\cal H}\to \zeta_3$:
{\sf Panel A} -- $\chi_{c0}$.
Legend.
Solid purple curve -- $x {\mathpzc c}^{\mathsf 0}(x;\zeta_3)$ obtained using mass-dependent splitting functions;
dot-dashed cyan -- $x {\mathpzc c}^{\mathsf 0}(x;\zeta_3)$ with mass-independent valence dof splitting;
dashed green -- $x {\mathpzc u}^{\pi}(x;\zeta_3)$;
dotted grey -- $(1/3)x{\mathpzc c}^{\mathsf 0}(x;\zeta_{\cal H})$.
{\sf Panel B} -- $\eta_c$.
Legend. Analogous to that in {\sf A}.
}
\end{figure}

\begin{figure}[t]
\vspace*{1ex}

\hspace*{-1ex}\begin{tabular}{l}
\centerline{\includegraphics[clip, width=0.44\textwidth]{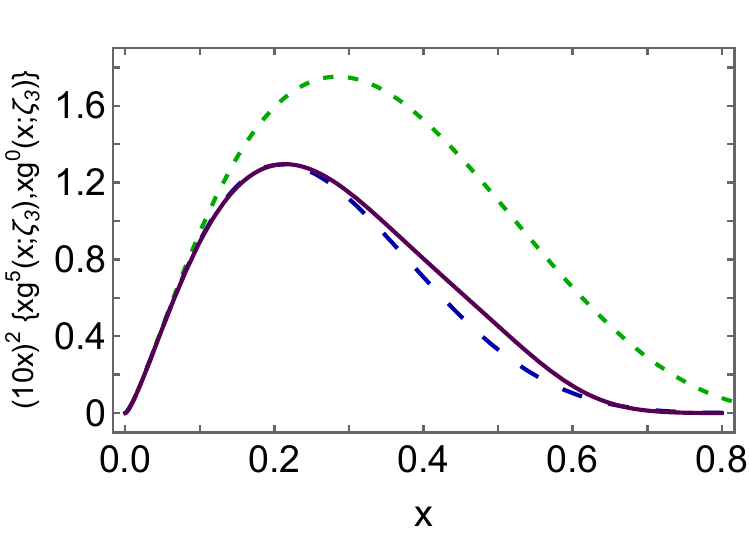}}\\[-37.5ex]
\normalsize{\textsf{glue}\mbox{\small $(\zeta_3)$} $\quad \mbox{\rm\small DF}$}
\end{tabular}
\vspace*{31ex}
\caption{\label{glueDFsevolved}
Glue DFs at $\zeta_3$ in the
$\chi_{c0}$ (solid purple),
$\eta_{c}$ (dashed blue),
and $\pi$ (short-dashed green).
}
\end{figure}

\section{Parton distribution functions: evolved}
\label{DFevolved}
Using the AO scheme with mass-dependent splitting functions for valence and singlet DFs \cite{Lu:2022cjx}, \cite[Sect.\,7.3]{Cui:2020tdf}, we evolve the hadron-scale DFs described in Sect.\,\ref{DFhadron} $\zeta_{\cal H} \to \zeta_3 :=  3.2\,$GeV, with the results depicted in Figs.\,\ref{DFsevolved}, \ref{glueDFsevolved}: nonzero glue and sea DFs are exposed by such evolution.
\emph{N.B}.\ Our AO implementation ensures that $c$, $\bar c$ valence dof do not emit as much glue as lighter valence dof \cite[Sect.\,7.3]{Cui:2020tdf}: such emission is suppressed for $\zeta \lesssim M_c$ and the impact of this is seen in the differences between solid and dot-dashed curves in Fig.\,\ref{DFsevolved}.

Regarding Fig.\,\ref{DFsevolved}A, the two-humped $\chi_{c0}$ DF structure is still perceptible at this scale, although the absolute minimum has become a diffuse region containing an inflection point.
In comparison with the $\pi$ valence DF, the $\chi_{c0}$ DF has far less support on $x\gtrsim 0.6$.
The same is true for the $\eta_{c}$ DF drawn in Fig.\,\ref{DFsevolved}B.
The images in Fig.\,\ref{DFsevolved} suggest that distinctions between scalar and pseudoscalar DFs disappear under evolution to increasingly larger scales.

Glue DFs are drawn in Fig.\,\ref{glueDFsevolved}.
Evidently, despite differences between the valence dof DFs, the $\chi_{c0}(\zeta_3)$ and $\eta_c(\zeta_3)$ glue DFs are practically indistinguishable.
Unsurprisingly, however, they are distinct from the in-pion glue DF.
This owes largely to the following two features:
hadron-scale $\chi_{c0}$, $\eta_c$ DFs are endpoint suppressed, Fig.\,\ref{CompareX0DFs};
and glue-parton emission by valence dof in $\chi_{c0}$, $\eta_c$ is damped by the heavier mass of the $c$ quark compared to that of the light quarks.
Qualitatively identical images can be plotted for the sea quark DFs; but since they reveal nothing essentially new, they are omitted herein.
(Analogous conclusions hold for DFs of the pion and its first radial excitation \cite{Xu:2025cyj}.)

Of course, $\chi_{c0}$ and $\eta_c$ targets, real or virtual, are unlikely to ever be achievable, so measurements that may be used to infer the DFs in Figs.\,\ref{DFsevolved}, \ref{glueDFsevolved} are improbable.
Nevertheless, in completing these calculations, we have delivered a single-framework unification of nucleon and many meson structure function predictions \cite{Cui:2020tdf, Lu:2022cjx, Xu:2025cyj}; hence, our results should serve as valuable benchmarks for comparable theory attempts to expose local and global differences between the structural features of hadrons.
For that reason, in Table~\ref{Tabmomfrac}, we list the species decomposed light-front momentum fractions produced by the profiles in Figs.\,\ref{DFsevolved}, \ref{glueDFsevolved} and the associated sea DFs.
Plainly, with the same value of the hadron scale characterising every strong-interaction bound-state, then the momentum fraction carried by each species is identical in all hadrons built from valence dof with the same current mass.
Differences only emerge when valence-dof current-mass effects are introduced into the splitting-functions of the evolution kernels \cite[Sect.\,7.3]{Cui:2020tdf}.
So, all in-$\chi_{c0}$ and in-$\eta_c$ momentum fractions are identical; but in the pion, after evolution, valence dof account for less of the momentum, and glue and sea account for more.

\begin{table}[t]
\caption{ \label{Tabmomfrac}
Species light-front momentum fractions at $\zeta=\zeta_3 := 3.2\,$GeV.
The third row provides a comparison with pion values \cite{Lu:2022cjx}.
Column 1 is $(c+\bar c)_{\rm valence}$ for $\chi_{c0}$ and $\eta_c$; and $(u+ \bar d)_{\rm valence}$ for $\pi$.
The final row records the relative difference between the momentum fractions in the light (L) and heavy (H) systems.}
\begin{tabular*}
{\hsize}
{
l@{\extracolsep{0ptplus1fil}}
|c@{\extracolsep{0ptplus1fil}}
c@{\extracolsep{0ptplus1fil}}
c@{\extracolsep{0ptplus1fil}}
c@{\extracolsep{0ptplus1fil}}
c@{\extracolsep{0ptplus1fil}}
c@{\extracolsep{0ptplus1fil}}}\hline\hline
\centering
 & valence & glue & $(u+\bar u)_{\rm sea}\ $ & $(d+\bar d)_{\rm sea}\ $& $(s+\bar s)_{\rm sea}\ $& $(c+\bar c)_{\rm sea}\ $ \\
$\chi_{c0}\ $ & $0.49\ $ & $0.40\ $ & $0.033\ $ & $0.033\ $ & $0.027\ $& $0.017\ $\\
$\eta_{c}\ $ & $0.49\ $ & $0.40\ $ & $0.033\ $ & $0.033\ $ & $0.027\ $& $0.017\ $\\\hline
$\pi \ $ & $0.44\ $ & $0.44\ $ & $0.038\ $ & $0.038\ $ & $0.031\ $& $0.019\ $\\\hline
\rule{0ex}{2.5ex}
\%$_{\rm  diff.}^{L/H}\ $ & $-10\ $ & $10\ $ & $15\ $ & $15\ $ & $15\ $& $12\ $
\\\hline\hline
\end{tabular*}
\end{table}

\section{Perspective}
Given that the charm quark current mass is larger than that of the proton, \emph{viz}.\ $m_c \gtrsim m_p$, it is often supposed that charmonia provide atom-like systems for which rigorously controlled nonrelativistic calculations can be realistic; hence, may be seen as providing a keen probe of heavy-quark QCD.
Against this background, herein, we used continuum Schwinger function methods (CSMs) to examine and elucidate the structural properties of ground-state scalar and pseudoscalar charmonia.
The analysis revealed and stressed the following points.

In QCD, charmonia are described by Poincar\'e covariant Bethe-Salpeter wave functions (BSWFs), which can be calculated using CSMs [Sect.\,\ref{SecBSWF}].
Regarding their rest-frame orbital angular momentum structure (OAM), charmonia wave functions are very complicated [Sect.\,\ref{Sec3}].
Consequently, the $\chi_{c0}$ cannot reliably be described by a quark-model-like $P$-wave-dominant wave function; likewise, no analogous $S$-wave-dominant wave function is valid for the $\eta_c$.

With such BSWFs in hand, one can supply predictions for the quasiparticle leading-twist distribution amplitudes (DAs) of these systems [Sect.\,\ref{secDA}]: $\varphi_{\mathsf 0}$ and $\varphi_{\mathsf 5}$.
Symmetries entail that the $\chi_{c0}$ DA, $\varphi_{\mathsf 0}$, possesses domains of balanced negative and positive support; and with $m_c \gtrsim m_p$, this DA is endpoint suppressed when compared with QCD's asymptotic meson DA, $\varphi_{\rm as}$.
On the other hand, in comparison with $\varphi_{\rm as}$, the $\eta_c$ DA is contracted.  It is also endpoint suppressed.

Such BSWFs also enable predictions to be made for hadron-scale distribution functions associated with $\chi_{c0}$, $\eta_c$ valence degrees-of-freedom (dof): ${\mathpzc c}^{\mathsf 0}(\zeta_{\cal H})$ and ${\mathpzc c}^{\mathsf 5}(\zeta_{\cal H})$, respectively.
In completing the calculations [Sect.\,\ref{DFhadron}], we found that, to an excellent degree of approximation,
${\mathpzc c}^{\mathsf 0}(\zeta_{\cal H})\propto \varphi_{\mathsf 0}^2$;
and to a good but lesser degree of reliability,
${\mathpzc c}^{\mathsf 5}(\zeta_{\cal H})\propto \varphi_{\mathsf 5}^2$.
These outcomes signal that a separable approximation should be practically reliable for the light-front wave functions of these mesons.
Work is underway to quantify these qualitative remarks.

We subsequently evolved these hadron scale DFs to $\zeta = \zeta_3 := 3.2\,$GeV [Sect.\,\ref{DFevolved}].
Pointwise differences between ${\mathpzc c}^{\mathsf 0}(x;\zeta_{\cal H})$ and ${\mathpzc c}^{\mathsf 5}(x;\zeta_{\cal H})$ diminish under evolution.
Furthermore, glue and sea DFs emerge: they are practically the same in both systems.
Highlighting these things, species-separated light-front momentum fractions are identical in the $\chi_{c0}$ and $\eta_c$.
On the other hand, compared with analogous in-pion quantities, there are marked differences, \emph{e.g}., the in-pion glue momentum fraction is 10\% larger than that in the charmonia.

Our study suggests that charmonia are more complex systems than is commonly imagined and that care should be taken when attempting to draw connections between the properties of such systems and heavy-quark QCD.

\medskip

\noindent\textbf{Acknowledgments}.
We thank K.\ Raya and Z.-Q.\ Yao for constructive comments.
Work supported by:
National Natural Science Foundation of China, grant no.\ 12135007;
and
Ministerio Espa\~nol de Ciencia, Innovaci\'on y Universidades (MICINN) grant no.\ PID2022-140440NB-C22.

\medskip
\noindent\textbf{Data Availability Statement}. This manuscript has no associated data or the data will not be deposited. [Authors' comment: All information necessary to reproduce the results described herein is contained in the material presented above.]

\medskip
\noindent\textbf{Declaration of Competing Interest}.
The authors declare that they have no known competing financial interests or personal relationships that could have appeared to influence the work reported in this paper.


\end{document}